\documentstyle[osa,manuscript]{revtex}

\input epsf.sty 
\input epsfig.sty

\begin{document}

\thispagestyle{empty}
\renewcommand{\thefootnote}{\fnsymbol{footnote}}


{\*\vspace{2.0cm}} 

\begin{center}
{\bf\Large Hamiltonian Formalism for Solving the Vlasov-Poisson 
Equations and Its Applications to Periodic Focusing Systems and 
the Coherent Beam-Beam Interaction} 

\vspace{1cm} 

Stephan I. Tzenov and Ronald C. Davidson\\ 
{\it Plasma Physics Laboratory, Princeton University, Princeton, 
New Jersey 08543} 
\end{center} 

\medskip 

\vfill

\renewcommand{\baselinestretch}{1}
\normalsize

\begin{center}
{\bf\large   
Abstract }
\end{center}

\begin{quote}
A Hamiltonian approach to the solution of the Vlasov-Poisson equations 
has been developed. Based on a nonlinear canonical transformation, the 
rapidly oscillating terms in the original Hamiltonian are transformed 
away, yielding a new Hamiltonian that contains slowly varying terms 
only. The formalism has been applied to the dynamics of an intense 
beam propagating through a periodic focusing lattice, and to the 
coherent beam-beam interaction. A stationary solution to the 
transformed Vlasov equation has been obtained. 
\end{quote}

\vfill


\begin{center} 
{\it Submitted to the} {\bf Physical Review Special Topics -- 
Accelerators and Beams} 
\end{center}

\newpage


\renewcommand{\baselinestretch}{1}
\normalsize


%
\pagestyle{plain}

\section{INTRODUCTION}

The evolution of charged particle beams in accelerators and storage 
rings can often be described by the Vlasov-Maxwell equations 
\cite{davidson,chaobook}. At high energies the discrete-particle 
collision term \cite{tzenov1} comprises a small correction to the 
dynamics and can be neglected. Radiation effects at sufficiently high 
energies for leptons can be a significant feature of the dynamics, 
and should normally be included in the model under consideration.

The Vlasov-Maxwell equations constitute a considerable simplification 
in the description of charged particle beam propagation. Nonetheless 
there are only a few cases that are tractable analytically 
\cite{davidson,chaobook}. Therefore, it is of the utmost importance to 
develop a systematic perturbation approach, able to provide satisfactory 
results in a wide variety of cases of physical interest. 

Particle beams are subject to external forces that are often rapidly 
oscillating, such as quadrupolar focusing forces, RF fields, etc. In 
addition, the collective self-field excitations can be rapidly 
oscillating as well. A typical example is a colliding-beam storage 
ring device, where the evolution of each beam is strongly affected 
by the electromagnetic force produced by the counter-propagating beam 
\cite{cai,chao,seeman}. The beam-beam kick each beam experiences is 
localized only in a small region around the interaction point, and is 
periodic with a period of one turn. 

In this and other important applications, one is primarily interested in 
the long-time behavior of the beam, thus discarding the fast processes 
on time scales of order the period of the rapid oscillations. To extract 
the relevant information, an efficient method of averaging is developed 
in Sec. 2. Unlike the standard canonical perturbation technique 
\cite{channell,qin}, the approach used here is carried out in a ``mixed'' 
phase space (old coordinates and new canonical momenta) \cite{tzenov2}, 
which is simpler and more efficient in a computational sense. It should 
be pointed out that the formalism developed here is strictly speaking 
non-canonical and in general does not provide complete elimination of 
fast oscillating terms in the transformed Vlasov equation in the mixed 
variables. Nevertheless, such an elimination can be performed in the 
new Hamiltonian in the mixed variables. Moreover, if the distribution 
function is assumed to be an arbitrary function of the new 
time-independent Hamiltonian, it is a stationary solution of the 
nonlinear Vlasov equation in the mixed variables. The canonical 
perturbation method developed in this paper is further applied to 
intense beam propagation in a periodic focusing structure (Sec. 3) 
and to the coherent beam-beam interaction (Secs. 4 and 5). A coupled 
set of nonlinear integral equations for the equilibrium beam densities 
is derived. 

To summarize, the effectiveness of the Hamiltonian formalism developed 
in the present paper is demonstrated in two particular examples. In the 
first example discussed in Sec. 3, the short-scale dynamics is contained 
in the external focusing force acting on the beam, while an essential 
feature of the coherent beam-beam interaction treated in Secs. 4 and 5 
is the relatively fast variation of the collective interaction between 
the colliding beams. The simplicity in applying the Hamiltonian 
averaging technique is embedded in the use of mixed canonical variables. 
Instead of expanding the generating function and the new Hamiltonian 
in terms of the new canonical coordinates and momenta \cite{channell,qin} 
one has to simply solve the Hamilton-Jacobi equations order by order. 
It should be emphasized that the mixed variable Hamiltonian formalism 
can be used to derive amplitude equations, describing processes of 
formation of patterns and coherent structures in a number of plasma and 
beam systems in which collective processes are important.

\section{THE HAMILTONIAN FORMALISM} 

We consider a $N$-dimensional dynamical system, described by the 
canonical conjugate pair of vector variables 
${\left( {\bf q}, {\bf p} \right)}$ with components 
\begin{eqnarray}\label{eq:components} 
{\bf q} = {\left( q_1, q_2, \dots, q_N \right)}, 
\nonumber \\ 
{\bf p} = {\left( p_1, p_2, \dots, p_N \right)}. 
\end{eqnarray} 
\noindent 
The Vlasov equation for the distribution function 
$f{\left( {\bf q}, {\bf p}; t \right)}$ can be expressed as 
\begin{equation}\label{eq:vlasov} 
{\frac {\partial f} {\partial t}} + 
{\left[ f, H \right]}_{{\bf q}, {\bf p}} = 0, 
\end{equation} 
\noindent 
where 
\begin{equation}\label{eq:bracket} 
{\left[ F, G \right]}_{{\bf q}, {\bf p}} = 
{\frac {\partial F} {\partial q_i}} 
{\frac {\partial G} {\partial p_i}} - 
{\frac {\partial F} {\partial p_i}} 
{\frac {\partial G} {\partial q_i}} 
\end{equation} 
\noindent 
is the Poisson bracket, $H{\left( {\bf q}, {\bf p}; t \right)}$ is 
the Hamiltonian of the system, and summation over repeated indices is 
implied. Next we define a canonical transformation via the generating 
function of the second type according to 
\begin{equation}\label{eq:canonical} 
S = S {\left( {\bf q}, {\bf P}; t \right)}, 
\end{equation} 
\noindent 
and assume that the Hessian matrix 
\begin{equation}\label{eq:jacobian} 
{\widehat{\cal H}}_{ij} 
{\left( {\bf q}, {\bf P}; t \right)} = 
{\frac {\partial^2 S} {\partial q_i \partial P_j}} 
\end{equation} 
\noindent 
of the generating function $S {\left( {\bf q}, {\bf P}; t \right)}$ 
is non-degenerate, i.e., 
\begin{equation}\label{eq:degenerate} 
\det {\left( {\widehat{\cal H}}_{ij} \right)} \neq 0. 
\end{equation} 
\noindent 
This implies that the inverse matrix ${\widehat{\cal H}}_{ij}^{-1}$ 
exists. The new canonical variables ${\left( {\bf Q}, {\bf P} \right)}$ 
are defined by the canonical transformation as 
\begin{equation}\label{eq:transform} 
p_i = {\frac {\partial S} {\partial q_i}}, 
\qquad \qquad 
Q_i = {\frac {\partial S} {\partial P_i}}. 
\end{equation} 

We also introduce the distribution function defined in terms of the new 
canonical coordinates ${\left( {\bf Q}, {\bf P} \right)}$ and the 
mixed pair of canonical variables ${\left( {\bf q}, {\bf P} \right)}$ 
according to 
\begin{equation}\label{eq:distrib1} 
f_0 {\left( {\bf Q}, {\bf P}; t \right)} = 
f {\left( {\bf q} {\left( {\bf Q}, {\bf P}; t \right)}, 
{\bf p} {\left( {\bf Q}, {\bf P}; t \right)}; t \right)}, 
\end{equation} 
\begin{equation}\label{eq:distrib2} 
F_0{\left( {\bf q}, {\bf P}; t \right)} = 
f{\left( {\bf q}, {\bf p} 
{\left( {\bf q}, {\bf P}; t \right)}; t \right)}. 
\end{equation} 
\noindent 
In particular, in Eq. (\ref{eq:distrib1}) the old canonical variables 
are expressed in terms of the new ones, which is ensured by the 
implicit function theorem, provided the relation (\ref{eq:degenerate}) 
holds. As far as the function $F_0{\left( {\bf q}, {\bf P}; t \right)}$ 
is concerned, we simply replace the old momentum ${\bf p}$ by its 
counterpart taken from the first of Eqs. (\ref{eq:transform}). Because 
\begin{equation}\label{eq:intermed1} 
{\frac {\partial p_i} {\partial P_j}} = 
{\frac {\partial^2 S} {\partial q_i \partial P_j}} = 
{\widehat{\cal H}}_{ij} \qquad \Longrightarrow \qquad 
{\frac {\partial P_i} {\partial p_j}} = 
{\widehat{\cal H}}_{ij}^{-1}, 
\end{equation} 
\noindent 
we can express the Poisson bracket in terms of the mixed variables 
in the form 
\begin{equation}\label{eq:bracketmix} 
{\left[ F, G \right]}_{{\bf q}, {\bf P}} = 
{\widehat{\cal H}}_{ji}^{-1} {\left( 
{\frac {\partial F} {\partial q_i}} 
{\frac {\partial G} {\partial P_j}} - 
{\frac {\partial F} {\partial P_j}} 
{\frac {\partial G} {\partial q_i}} 
\right)}. 
\end{equation} 
\noindent 
Differentiation of Eq. (\ref{eq:transform}) with respect to time 
$t$, keeping the old variables ${\left( {\bf q}, {\bf p} \right)}$ 
fixed, yields 
\begin{equation}\label{eq:intermed2} 
{\frac {\partial^2 S} {\partial q_i \partial t}} + 
{\frac {\partial^2 S} {\partial q_i \partial P_j}} 
{\left( {\frac {\partial P_j} {\partial t}} 
\right)}_{qp} = 0, 
\end{equation} 
\begin{equation}\label{eq:intermed3} 
{\left( {\frac {\partial Q_i} {\partial t}} 
\right)}_{qp} = 
{\frac {\partial^2 S} {\partial P_i \partial t}} + 
{\frac {\partial^2 S} {\partial P_i \partial P_j}} 
{\left( {\frac {\partial P_j} {\partial t}} 
\right)}_{qp}, 
\end{equation} 
\noindent 
or 
\begin{equation}\label{eq:intermed4} 
{\left( {\frac {\partial P_j} {\partial t}} 
\right)}_{qp} = - {\widehat{\cal H}}_{ji}^{-1} 
{\frac {\partial^2 S} {\partial q_i \partial t}}. 
\end{equation} 

Our goal is to express the Vlasov equation (\ref{eq:vlasov}) in terms 
of the mixed variables ${\left( {\bf q}, {\bf P} \right)}$. Taking 
into account the identities 
\begin{equation}\label{eq:intermed5} 
{\frac {\partial Q_i} {\partial q_j}} = 
{\frac {\partial^2 S} {\partial q_j \partial P_i}} = 
{\widehat{\cal H}}_{ji} \quad \Longrightarrow \quad 
{\frac {\partial q_i} {\partial Q_j}} = 
{\widehat{\cal H}}_{ji}^{-1}, 
\end{equation} 
\begin{equation}\label{eq:intermed6} 
{\frac {\partial f_0} {\partial Q_i}} = 
{\widehat{\cal H}}_{ij}^{-1} 
{\frac {\partial F_0} {\partial q_j}}, 
\end{equation} 
\begin{equation}\label{eq:intermed7} 
{\frac {\partial f_0} {\partial P_i}} = 
{\frac {\partial F_0} {\partial P_i}} - 
{\frac {\partial f_0} {\partial Q_j}} 
{\frac {\partial^2 S} {\partial P_i \partial P_j}}, 
\end{equation} 
\noindent 
we obtain 
\begin{eqnarray} 
{\left( {\frac {\partial f} {\partial t}} 
\right)}_{qp} = 
{\frac {\partial f_0} {\partial t}} + 
{\frac {\partial f_0} {\partial Q_i}} 
{\left( {\frac {\partial Q_i} {\partial t}} 
\right)}_{qp} + 
{\frac {\partial f_0} {\partial P_i}} 
{\left( {\frac {\partial P_i} {\partial t}} 
\right)}_{qp} \nonumber 
\end{eqnarray} 
\begin{eqnarray} 
= {\frac {\partial F_0} {\partial t}} + 
{\widehat{\cal H}}_{ji}^{-1} {\left( 
{\frac {\partial F_0} {\partial q_i}} 
{\frac {\partial^2 S} {\partial t \partial P_j}} - 
{\frac {\partial F_0} {\partial P_j}} 
{\frac {\partial^2 S} {\partial t \partial q_i}} 
\right)} \nonumber 
\end{eqnarray} 
\begin{equation}\label{eq:vlasovmix} 
= {\frac {\partial F_0} {\partial t}} + 
{\left[ F_0, {\frac {\partial S} {\partial t}} 
\right]}_{{\bf q}, {\bf P}}. 
\end{equation} 
\noindent 
Furthermore, using the relation 
\begin{equation}\label{eq:relation} 
{\left[ f, H \right]}_{{\bf q}, {\bf p}} = 
{\left[ F_0, {\cal H} \right]}_{{\bf q}, {\bf P}}, 
\end{equation} 
\noindent 
where 
\begin{equation}\label{eq:newhamil} 
{\cal H} {\left( {\bf q}, {\bf P}; t \right)} = 
H {\left( {\bf q}, {\bf \nabla}_q S; t \right)}, 
\end{equation} 
\noindent 
we express the Vlasov equation in terms of the mixed variables 
according to 
\begin{equation}\label{eq:mixvlasov} 
{\frac {\partial F_0} {\partial t}} + 
{\left[ F_0, {\cal K} \right]}_{{\bf q}, {\bf P}} = 0, 
\end{equation} 
\noindent 
where
\begin{equation}\label{eq:newhamilt} 
{\cal K} {\left( {\bf q}, {\bf P}; t \right)} = 
{\frac {\partial S} {\partial t}} + 
H {\left( {\bf q}, {\bf \nabla}_q S; t \right)} 
\end{equation} 
\noindent 
is the new Hamiltonian. 

For the distribution function 
$f_0 {\left( {\bf Q}, {\bf P}; t \right)}$, depending on the new 
canonical variables, we clearly obtain 
\begin{equation}\label{eq:equilib} 
{\frac {\partial f_0} {\partial t}} + 
{\left[ f_0, {\cal K} \right]}_{{\bf Q}, {\bf P}} = 0, 
\end{equation} 
\noindent 
where the new Hamiltonian ${\cal K}$ is a function of the new 
canonical pair ${\left( {\bf Q}, {\bf P} \right)}$, such that 
\begin{equation}\label{eq:newhamilto} 
{\cal K} {\left( {\bf \nabla}_P S, {\bf P}; t \right)} = 
{\frac {\partial S} {\partial t}} + 
H {\left( {\bf q}, {\bf \nabla}_q S; t \right)}, 
\end{equation} 
\noindent 
and the Poisson bracket entering Eq. (\ref{eq:equilib}) has the same 
form as Eq. (\ref{eq:bracket}), expressed in the new canonical 
variables.

\section{PROPAGATION OF AN INTENSE BEAM THROUGH A PERIODIC FOCUSING 
LATTICE} 

As a first application of the Hamiltonian formalism, we consider the 
propagation of a continuous beam through a periodic focusing lattice 
in a circular ring with radius $R$. Particle motion is accomplished 
in two degrees of freedom in a plane transverse to the design orbit. 
The model equations consist of the nonlinear Vlasov-Poisson equations 
\cite{davidson} 
\begin{equation}\label{eq:vlasove} 
{\frac {\partial f} {\partial \theta}} + 
{\left[ f, H \right]}_{{\bf q}, {\bf p}} = 0, 
\end{equation} 
\begin{equation}\label{eq:poissone} 
{\nabla}_{\bf q}^2 \psi = - 4 \pi \varrho = 
- 4 \pi \! \int \! d^2 {\bf p} f 
{\left( {\bf q}, {\bf p}; \theta \right)}, 
\end{equation} 
\noindent 
where 
\begin{equation}\label{eq:hamiltoniane} 
H {\left( {\bf q}, {\bf p}; \theta \right)} = 
{\frac {R} {2}} {\left( p_x^2 + p_z^2 \right)} + 
{\frac {1} {2R}} {\left( G_x x^2 + 
G_z z^2 \right)} + \lambda \psi 
{\left( {\bf q}; \theta \right)} 
\end{equation} 
\noindent 
is the normalized Hamiltonian, and ${\bf q} = (x, z)$. The transverse 
canonical momenta ${\bf p} = {\left( p_x, p_z \right)}$ entering the 
Hamiltonian (\ref{eq:hamiltoniane}) are dimensionless variables which 
represent the actual transverse momenta of the moving particle scaled 
by the longitudinal momentum of the synchronous particle 
\cite{tzenovbook}. 

In addition, $R$ is the mean radius of the accelerator and $\psi$ is 
a normalized potential related to the actual electric potential 
$\varphi$ according to 
\begin{equation}\label{eq:normpot} 
\psi = {\frac {4 \pi \varepsilon_0} {N e_b}} 
\varphi, 
\end{equation} 
\noindent 
where $N$ is the total number of particles in the beam, $e_b$ is the 
particle charge, and $\varepsilon_0$ is the electric susceptibility of 
vacuum. Moreover, the parameter $\lambda$ is defined by 
\begin{equation}\label{eq:lamparam} 
\lambda = {\frac {NR r_b} 
{\beta_s^2 \gamma_s^3}}, 
\end{equation} 
\noindent 
where $\beta_s = v_s / c$ is the relative velocity of the synchronous 
particle, $\gamma_s = {\left( 1 - \beta_s^2 \right)}^{- 1/2}$ is 
the Lorentz factor, and 
\begin{equation}\label{eq:elradius} 
r_b = {\frac {e_b^2} 
{4 \pi \varepsilon_0 m_b c^2}} 
\end{equation} 
\noindent 
is the classical radius of a beam particle with charge $e_b$ and rest 
mass $m_b$. The coefficients $G_{x,z} {\left( \theta \right)}$ 
determining the focusing strength in both transverse directions are 
periodic functions of $\theta$ 
\begin{equation}\label{eq:focoeff} 
G_{x,z} {\left( \theta + \Theta \right)} = 
G_{x,z} {\left( \theta \right)}, 
\end{equation} 
\noindent 
with period $\Theta$. 

Following the procedure outlined in the preceding section we transform 
Eqs. (\ref{eq:vlasove}) -- (\ref{eq:hamiltoniane}) according to 
\begin{equation}\label{eq:vlasovst} 
{\left[ F_0, {\cal K} \right]}_{{\bf q}, {\bf P}} 
\equiv 0, 
\end{equation} 
\begin{equation}\label{eq:hamilst} 
{\frac {\partial S} {\partial \theta}} + 
\epsilon H {\left( {\bf q}, {\nabla}_{\bf q} S; 
\theta \right)} = 
{\cal K} {\left( {\bf q}, {\bf P} \right)}, 
\end{equation} 
\begin{equation}\label{eq:poisst} 
{\nabla}_{\bf q}^2 \psi = - 4 \pi \! \int \! 
d^2 {\bf P} F_0 {\left( {\bf q}, {\bf P} \right)} 
\det {\left( {\nabla}_{\bf q} 
{\nabla}_{\bf P} S \right)}, 
\end{equation} 
\noindent 
where $\epsilon$ is formally a small parameter proportional to the 
focusing field strength, which will be set equal to unity at the end 
of the calculation. The next step is to expand the quantities $S$, 
${\cal K}$ and $\psi$ in a power series in $\epsilon$ according to 
\begin{equation}\label{eq:expandsf} 
S = S_0 + \epsilon S_1 + \epsilon^2 S_2 + 
\epsilon^3 S_3 + \dots, 
\end{equation} 
\begin{equation}\label{eq:expandkf} 
{\cal K} = {\cal K}_0 + \epsilon {\cal K}_1 + 
\epsilon^2 {\cal K}_2 + 
\epsilon^3 {\cal K}_3 + \dots, 
\end{equation} 
\begin{equation}\label{eq:expandphi} 
\psi = \psi_0 + \epsilon \psi_1 + 
\epsilon^2 \psi_2 + 
\epsilon^3 \psi_3 + \dots. 
\end{equation} 
\noindent 
We now substitute the expansions (\ref{eq:expandsf}) -- 
(\ref{eq:expandphi}) into Eqs. (\ref{eq:hamilst}) and 
(\ref{eq:poisst}) and obtain perturbation equations that can be 
solved order by order. 

The lowest order solution is evident and has the form 
\begin{equation}\label{eq:sols0} 
S_0 = {\bf q} \cdot {\bf P}, 
\qquad \qquad 
{\cal K}_0 \equiv 0, 
\end{equation} 
\begin{equation}\label{eq:solphi0} 
{\nabla}_{\bf q}^2 \psi_0 =  - 4 \pi \! \int \! 
d^2 {\bf P} F_0 {\left( {\bf q}, {\bf P} \right)}. 
\end{equation} 

{\it First order} $O(\epsilon)$: Taking into account the already 
obtained lowest order solutions (\ref{eq:sols0}) and 
(\ref{eq:solphi0}), the Hamilton-Jacobi equation (\ref{eq:hamilst}) 
to first order in $\epsilon$ can be expressed as 
\begin{equation}\label{eq:hamjac1} 
{\frac {\partial S_1} {\partial \theta}} + 
{\frac {R} {2}} {\left( P_x^2 + P_z^2 \right)} + 
{\frac {1} {2R}} {\left( G_x x^2 + 
G_z z^2 \right)} + \lambda \psi_0 = 
{\cal K}_1 {\left( {\bf q}, {\bf P} \right)}. 
\end{equation} 
\noindent 
Imposing the condition that the first order Hamiltonian ${\cal K}_1$ 
be equal to 
\begin{equation}\label{eq:hamilt1} 
{\cal K}_1 {\left( {\bf q}, {\bf P} \right)} = 
{\frac {R} {2}} {\left( P_x^2 + P_z^2 \right)} + 
{\frac {1} {2R}} {\left( {\overline{G}}_x x^2 + 
{\overline{G}}_z z^2 \right)} + \lambda \psi_0 
{\left( {\bf q} \right)}, 
\end{equation} 
\noindent 
we obtain immediately 
\begin{equation}\label{eq:sfunc1} 
S_1 = - {\frac {1} {2R}} {\left[ 
{\widetilde{G}}_x (\theta) x^2 + 
{\widetilde{G}}_z (\theta) z^2 \right]}, 
\end{equation} 
\begin{equation}\label{eq:phi1} 
\psi_1 \equiv 0. 
\end{equation} 
Here we have introduced the notation 
\begin{equation}\label{eq:notation} 
{\overline{G}}_{x,z} = {\frac {1} {\Theta}} 
\int \limits_{\theta_0}^{\theta_0 + \Theta} 
d \theta G_{x,z} (\theta), 
\qquad \qquad 
{\widetilde{G}}_{x,z} (\theta) = 
\int \limits_{\theta_0}^{\theta_0 + \theta} 
d \tau {\left[ G_{x,z} (\tau) - 
{\overline{G}}_{x,z} \right]}. 
\end{equation} 

Note that since the focusing coefficients are periodic functions of 
$\theta$ they can be expanded in a Fourier series 
\begin{equation}\label{eq:fourier} 
G_{x,z} (\theta) = \sum 
\limits_{n=- \infty}^{\infty} G_{x,z}^{(n)} 
\exp{\left( in \Omega \theta \right)}, 
\end{equation} 
\noindent 
where 
\begin{equation}\label{eq:fourier1} 
G_{x,z}^{(n)} = {\frac {1} {\Theta}} 
\int \limits_0^{\Theta} d \theta 
G_{x,z} (\theta) \exp{\left( - in \Omega \theta 
\right)}, 
\end{equation} 
\noindent 
and $\Omega = 2 \pi / \Theta$. Therefore for the quantities 
${\overline{G}}_{x,z}$ and ${\widetilde{G}}_{x,z} (\theta)$ expressed 
in terms of the Fourier amplitudes, we obtain 
\begin{equation}\label{eq:fourampl} 
{\overline{G}}_{x,z} = G_{x,z}^{(0)}, 
\qquad \qquad \qquad 
{\widetilde{G}}_{x,z} (\theta) = - 
{\frac {i} {\Omega}} \sum 
\limits_{n \neq 0} {\frac {G_{x,z}^{(n)}} {n}} 
\exp{\left( in \Omega \theta \right)}. 
\end{equation} 

{\it Second order} $O(\epsilon^2)$: To this order, the 
Hamilton-Jacobi equation (\ref{eq:hamilst}) takes the form 
\begin{equation}\label{eq:hamjac2} 
{\frac {\partial S_2} {\partial \theta}} - 
{\left( x P_x {\widetilde{G}}_x + 
z P_z {\widetilde{G}}_z \right)} = 
{\cal K}_2 {\left( {\bf q}, {\bf P} \right)}. 
\end{equation} 
\noindent 
It is straightforward to solve Eq. (\ref{eq:hamjac2}), yielding the 
obvious result 
\begin{equation}\label{eq:sfunc2} 
S_2 = x P_x {\widetilde{\widetilde{G}}}_x 
(\theta) + z P_z {\widetilde{\widetilde{G}}}_z 
(\theta), \qquad \qquad \qquad 
{\cal K}_2 {\left( {\bf q}, {\bf P} \right)} 
\equiv 0. 
\end{equation} 
\noindent 
For the second order potential $\psi_2$ we obtain the equation 
\begin{equation}\label{eq:phipoiss2} 
{\nabla}_{\bf q}^2 \psi_2 =  - 4 \pi 
{\left( {\widetilde{\widetilde{G}}}_x + 
{\widetilde{\widetilde{G}}}_z \right)} 
\! \int \! d^2 {\bf P} 
F_0 {\left( {\bf q}, {\bf P} \right)}, 
\end{equation} 
\noindent 
or, making use of (\ref{eq:solphi0}), 
\begin{equation}\label{eq:phi2} 
\psi_2 {\left( {\bf q}; \theta \right)} = 
{\left[ {\widetilde{\widetilde{G}}}_x (\theta) + 
{\widetilde{\widetilde{G}}}_z (\theta) \right]} 
\psi_0 {\left( {\bf q} \right)}. 
\end{equation} 
\noindent 
In Eqs. (\ref{eq:sfunc2}) -- (\ref{eq:phi2}), 
${\widetilde{\widetilde{G}}}_{x,z} (\theta)$ denotes application of 
the integral operation in Eq. (\ref{eq:notation}) to 
${\widetilde{G}}_{x,z} (\theta)$, i.e., 
\begin{equation}\label{eq:tilnotat} 
{\widetilde{\widetilde{G}}}_{x,z} (\theta) = 
\int \limits_{\theta_0}^{\theta_0 + \theta} 
d \tau {\widetilde{G}}_{x,z} (\tau), 
\end{equation} 
\noindent 
because ${\overline{\widetilde{G}}}_{x,z} = 0$. 

{\it Third order} $O(\epsilon^3)$: To third order in $\epsilon$, 
the Hamilton-Jacobi equation (\ref{eq:hamilst}) can be written as 
\begin{equation}\label{eq:hamjac3} 
{\frac {\partial S_3} {\partial \theta}} + 
R {\left( P_x^2 {\widetilde{\widetilde{G}}}_x + 
P_z^2 {\widetilde{\widetilde{G}}}_z 
\right)} + {\frac {1} {2R}} {\left( 
{\widetilde{G}}_x^2 x^2 + 
{\widetilde{G}}_z^2 z^2 \right)} + \lambda 
{\left( {\widetilde{\widetilde{G}}}_x + 
{\widetilde{\widetilde{G}}}_z \right)} 
\psi_0 = 
{\cal K}_3 {\left( {\bf q}, {\bf P} \right)}. 
\end{equation} 
\noindent 
The third-order Hamiltonian ${\cal K}_3$ is given by the expression 
\begin{equation}\label{eq:hamilt3} 
{\cal K}_3 {\left( {\bf q}, {\bf P} \right)} = 
{\frac {1} {2R}} 
{\left( {\overline{{\widetilde{G}}_x^2}} x^2 + 
{\overline{{\widetilde{G}}_z^2}} z^2  \right)}. 
\end{equation} 
\noindent 
Equation (\ref{eq:hamjac3}) can be easily solved for the third-order 
generating function $S_3$. The result is 
\begin{equation}\label{eq:sfunc3} 
S_3 = - R {\left( P_x^2 
{\widetilde{\widetilde{\widetilde{G}}}}_x + 
 P_z^2 {\widetilde{\widetilde{\widetilde{G}}}}_z
\right)} - {\frac {1} {2R}} 
{\left( {\widetilde{{\widetilde{G}}_x^2}} x^2 + 
{\widetilde{{\widetilde{G}}_z^2}} z^2  \right)} - 
\lambda {\left( 
{\widetilde{\widetilde{\widetilde{G}}}}_x + 
{\widetilde{\widetilde{\widetilde{G}}}}_z \right)} 
\psi_0. 
\end{equation} 
\noindent 
For the third-order electric potential $\psi_3$ we obtain simply 
\begin{equation}\label{eq:phi3} 
\psi_3 \equiv 0. 
\end{equation} 

{\it Fourth order} $O(\epsilon^4)$: To the fourth order in the 
expansion parameter $\epsilon$ the Hamilton-Jacobi equation 
(\ref{eq:hamilst}) can be expressed as 
\begin{equation}\label{eq:hamjac4} 
{\frac {\partial S_4} {\partial \theta}} - 
x P_x {\left( {\widetilde{{\widetilde{G}}_x^2}} 
+ {\widetilde{G}}_x {\widetilde{\widetilde{G}}}_x 
\right)} - 
z P_z {\left( {\widetilde{{\widetilde{G}}_z^2}} 
+ {\widetilde{G}}_z {\widetilde{\widetilde{G}}}_z 
\right)} - \lambda R 
{\left( {\widetilde{\widetilde{G}}}_x + 
{\widetilde{\widetilde{G}}}_z \right)} 
{\left( P_x {\frac {\partial \psi_0} 
{\partial x}} +  P_z {\frac {\partial \psi_0} 
{\partial z}} \right)} = 
{\cal K}_4 {\left( {\bf q}, {\bf P} \right)}. 
\end{equation} 
\noindent 
The obvious condition to impose is that the fourth-order Hamiltonian 
${\cal K}_4$ be equal to 
\begin{equation}\label{eq:hamilt4} 
{\cal K}_4 {\left( {\bf q}, {\bf P} \right)} = 
- x P_x 
{\overline{{\widetilde{G}}_x {\widetilde{\widetilde{G}}}_x}} 
- z P_z 
{\overline{{\widetilde{G}}_z {\widetilde{\widetilde{G}}}_z}}. 
\end{equation} 
\noindent 
With Eq. (\ref{eq:hamilt4}) in hand, it is straightforward to solve the 
fourth-order Hamilton-Jacobi equation (\ref{eq:hamjac4}) for $S_4$. We 
obtain 
\begin{equation}\label{eq:sfunc4} 
S_4 = 
x P_x {\left( {\widetilde{\widetilde{{\widetilde{G}}_x^2}}} 
+ {\widetilde{{\widetilde{G}}_x {\widetilde{\widetilde{G}}}_x}} 
\right)} + 
z P_z {\left( {\widetilde{\widetilde{{\widetilde{G}}_z^2}}} 
+ {\widetilde{{\widetilde{G}}_z {\widetilde{\widetilde{G}}}_z}} 
\right)} + \lambda R 
{\left( {\widetilde{\widetilde{\widetilde{G}}}}_x + 
{\widetilde{\widetilde{\widetilde{G}}}}_z \right)} 
{\left( P_x {\frac {\partial \psi_0} 
{\partial x}} +  P_z {\frac {\partial \psi_0} 
{\partial z}} \right)}. 
\end{equation} 
\noindent 
For the fourth-order electric potential $\psi_4$, we obtain the 
Poisson equation 
\begin{equation}\label{eq:phipoiss4} 
{\nabla}_{\bf q}^2 \psi_4 = 
{\left[ {\widetilde{\widetilde{G}}}_x 
{\widetilde{\widetilde{G}}}_z + 
{\widetilde{\widetilde{{\widetilde{G}}_x^2}}} + 
{\widetilde{\widetilde{{\widetilde{G}}_z^2}}} + 
{\widetilde{{\widetilde{G}}_x {\widetilde{\widetilde{G}}}_x}} + 
{\widetilde{{\widetilde{G}}_z {\widetilde{\widetilde{G}}}_z}} + 
\lambda R {\left( {\widetilde{\widetilde{\widetilde{G}}}}_x + 
{\widetilde{\widetilde{\widetilde{G}}}}_z \right)} 
{\nabla}_{\bf q}^2 \psi_0 \right]} 
{\nabla}_{\bf q}^2 \psi_0. 
\end{equation} 

{\it Fifth order} $O(\epsilon^5)$: In fifth order, we are interested 
in the Hamiltonian ${\cal K}_5$. Omitting algebraic details we find 
\begin{eqnarray}\nonumber 
{\cal K}_5 {\left( {\bf q}, {\bf P} \right)} = 
{\frac {R} {2}} {\left( 
{\overline{{\widetilde{\widetilde{G}}}_x^2}} P_x^2 + 
{\overline{{\widetilde{\widetilde{G}}}_z^2}} P_z^2 
\right)} + {\frac {1} {R}} {\left( 
{\overline{{\widetilde{G}}_x 
{\widetilde{{\widetilde{G}}_x^2}}}} x^2 + 
{\overline{{\widetilde{G}}_z 
{\widetilde{{\widetilde{G}}_z^2}}}} z^2 
\right)} 
\end{eqnarray} 
\begin{equation}\label{eq:hamilt5} 
+ \lambda {\left[ 
{\overline{{\widetilde{G}}_x 
{\left( {\widetilde{\widetilde{\widetilde{G}}}}_x + 
{\widetilde{\widetilde{\widetilde{G}}}}_z \right)}}} 
x {\frac {\partial \psi_0} {\partial x}} + 
{\overline{{\widetilde{G}}_z 
{\left( {\widetilde{\widetilde{\widetilde{G}}}}_x + 
{\widetilde{\widetilde{\widetilde{G}}}}_z \right)}}} 
z {\frac {\partial \psi_0} {\partial z}} 
\right]}. 
\end{equation} 

In concluding this section, we collect terms up to fifth order in 
$\epsilon$ in the new Hamiltonian ${\cal K} = {\cal K}_0 + \epsilon 
{\cal K}_1 + \epsilon^2 {\cal K}_2 + \dots$ and set $\epsilon = 1$. 
This gives 
\begin{eqnarray}\nonumber 
{\cal K} {\left( {\bf q}, {\bf P} \right)} = 
\sum \limits_{u=(x, z)} {\left( 
{\frac {R {\cal A}_u} {2}} P_u^2 + {\cal B}_u 
u P_u + {\frac {{\cal C}_u} {2R}} u^2 
\right)} + \lambda \psi_0 
{\left( {\bf q} \right)} 
\end{eqnarray} 
\begin{equation}\label{eq:newhami} 
+ \lambda {\left[ {\overline{{\widetilde{G}}_x 
{\left( {\widetilde{\widetilde{\widetilde{G}}}}_x + 
{\widetilde{\widetilde{\widetilde{G}}}}_z \right)}}} 
x {\frac {\partial \psi_0} {\partial x}} + 
{\overline{{\widetilde{G}}_z 
{\left( {\widetilde{\widetilde{\widetilde{G}}}}_x + 
{\widetilde{\widetilde{\widetilde{G}}}}_z \right)}}} 
z {\frac {\partial \psi_0} {\partial z}} 
\right]}, 
\end{equation} 
\noindent 
where the coefficients ${\cal A}_u$, ${\cal B}_u$ and ${\cal C}_u$ 
are defined by the expressions 
\begin{equation}\label{eq:coeffa} 
{\cal A}_u = 1 + \epsilon^4 
{\overline{{\widetilde{\widetilde{G}}}_u^2}}, 
\end{equation} 
\begin{equation}\label{eq:coeffb} 
{\cal B}_u = - \epsilon^3 
{\overline{{\widetilde{G}}_u 
{\widetilde{\widetilde{G}}}_u}}, 
\end{equation} 
\begin{equation}\label{eq:coeffc} 
{\cal C}_u = {\overline{G}}_u + \epsilon^2 
{\overline{{\widetilde{G}}_u^2}} + 2 \epsilon^4 
{\overline{{\widetilde{G}}_u 
{\widetilde{{\widetilde{G}}_u^2}}}}. 
\end{equation} 
\noindent 
The Hamiltonian (\ref{eq:newhami}), neglecting the contribution from 
the self-field $\psi_0$, describes the unperturbed betatron oscillations 
in both horizontal and vertical directions. 

It is useful to compute the unperturbed betatron tunes $\nu_{x,z}$ in 
terms of averages over the focusing field-strengths. For a Hamiltonian 
system governed by a quadratic form in the canonical variables of the 
type in Eq. (\ref{eq:newhami}), it is well-known that the characteristic 
frequencies $\nu_{x,z}$ can be expressed as 
\begin{equation}\label{eq:frequen} 
\nu_u^2 = {\cal A}_u {\cal C}_u - 
{\cal B}_u^2, 
\qquad \qquad \qquad 
{\left( u = x, z \right)}. 
\end{equation} 
\noindent 
Keeping terms up to sixth order in the perturbation parameter 
$\epsilon$, we obtain 
\begin{equation}\label{eq:frequency} 
\nu_u^2 = {\overline{G}}_u + \epsilon^2 
{\overline{{\widetilde{G}}_u^2}} + 
\epsilon^4 {\left( {\overline{G}}_u 
{\overline{{\widetilde{\widetilde{G}}}_u^2}} 
+ 2 {\overline{{\widetilde{G}}_u 
{\widetilde{{\widetilde{G}}_u^2}}}} \right)} 
+ \epsilon^6 {\left[ 
{\overline{{\widetilde{G}}_u^2}} 
{\overline{{\widetilde{\widetilde{G}}}_u^2}} 
- {\left( {\overline{{\widetilde{G}}_u 
{\widetilde{\widetilde{G}}}_u}} \right)}^2
\right]}. 
\end{equation} 
\noindent 
In terms of Fourier amplitudes of the focusing coefficients, the Eq. 
(\ref{eq:frequency}) can be expressed as 
\begin{eqnarray}\nonumber 
\nu_u^2 = G_u^{(0)} + {\frac {2 \epsilon^2} 
{\Omega^2}} \sum \limits_{n=1}^{\infty} 
{\frac {{\left| G_u^{(n)} \right|}^2} {n^2}} 
+ {\frac {2 \epsilon^4} {\Omega^4}} {\left[ 
G_u^{(0)} \sum \limits_{n=1}^{\infty} 
{\frac {{\left| G_u^{(n)} \right|}^2} {n^4}} 
+ 2 \sum \limits_{m,n=1 \atop m \neq n}^{\infty} 
{\frac {{\rm Re} {\left( G_u^{(m) \ast} 
G_u^{(n)} G_u^{(m-n)} \right)}} {mn (m-n)^2}} 
\right.} 
\end{eqnarray} 
\begin{equation}\label{eq:freqfourier} 
{\left. - 2 \sum \limits_{m,n=1}^{\infty} 
{\frac {{\rm Re} {\left( G_u^{(m)} G_u^{(n)} 
G_u^{(m+n) \ast} \right)}} {mn (m+n)^2}} 
\right]} + {\frac {4 \epsilon^6} 
{\Omega^6}} \sum \limits_{m, n=1}^{\infty} 
{\frac {{\left| G_u^{(m)} \right|}^2 
{\left| G_u^{(n)} \right|}^2} {m^2 n^4}}. 
\end{equation} 

For purposes of illustration, we consider a simple $FODO$ lattice with 
equal focusing and defocusing strengths $+G$ and $-G$, and period 
$\Theta$. We also assume that the longitudinal dimensions $\theta_f$ 
of the focusing and defocusing lenses are equal; the longitudinal 
dimensions $\theta_d$ of the corresponding drift spaces are assumed 
to be equal as well. Moreover, 
\begin{equation}\label{eq:dimperiod} 
2 {\left( \theta_f + 
\theta_d \right)} = \Theta. 
\end{equation} 
\noindent 
For simplicity we consider the horizontal degree of freedom only 
(the vertical one can be treated in analogous manner). The Fourier 
amplitudes of the focusing coefficients are 
\begin{equation}\label{eq:fodofourier} 
G_x^{(2n+1)} = {\frac {i G} {(2n+1) \pi}} 
{\left\{ \exp{\left[ -i (2n+1) \Omega 
\theta_f \right]} - 1 \right\}}, 
\qquad \qquad 
G_x^{(2n)} = 0, 
\end{equation} 
\noindent 
where $n=0, 1, 2, \dots$. To second order in $\epsilon$, we obtain for 
the horizontal betatron tune 
\begin{equation}\label{eq:fodotune} 
\nu_x^2 = {\frac {2 \epsilon^2 \Theta^2 G^2} 
{\pi^4}} \sum \limits_{m=1}^{\infty} 
{\frac {1} {(2m-1)^4}} 
\sin^2 {\frac {(2m-1) \pi \theta_f} {\Theta}}. 
\end{equation} 
\noindent 
In the limit of infinitely thin lenses, $\theta_f \rightarrow 0$, Eq. 
(\ref{eq:fodotune}) reduces to the well-known expression 
\begin{equation}\label{eq:fodothin} 
\nu_x^2 = {\frac {\epsilon^2 
\theta_f^2 G^2} {4}}, 
\end{equation} 
\noindent 
where use of the identity 
\begin{equation}\label{eq:fodoident} 
\sum \limits_{m=1}^{\infty} 
{\frac {1} {(2m-1)^2}} = 
{\frac {\pi^2} {8}} 
\end{equation} 
\noindent 
has been made. 

It is evident from Eqs. (\ref{eq:freqfourier}) and (\ref{eq:fodotune}), 
that the Hamiltonian averaging technique developed here represents a 
powerful formalism for evaluating the betatron tunes in terms of 
averages over the focusing field strength.

\section{COHERENT BEAM-BEAM INTERACTION} 

As a second application of the Hamiltonian formalism developed in 
Sec. 2, we study here the evolution of two counter-propagating 
beams, nonlinearly coupled by the electromagnetic interaction between 
the beams at collision. For simplicity, we consider one-dimensional 
motion in the vertical $(q)$ direction, described by the nonlinear 
Vlasov-Poisson equations 
\begin{equation}\label{eq:vlasovk} 
{\frac {\partial f_k} {\partial \theta}} + 
{\left[ f_k, H_k \right]} = 0, 
\end{equation} 
\begin{equation}\label{eq:poisson} 
{\frac {\partial^2 V_k} {\partial q^2}} = 
4 \pi \! \int \! dp f_{3-k} (q, p; \theta), 
\end{equation} 
\noindent 
where 
\begin{equation}\label{eq:hamiltonian} 
H_k = {\frac {\nu_k} {2}} 
{\left( p^2 + q^2 \right)} + 
\lambda_k \delta_p (\theta) 
V_k (q; \theta) 
\end{equation} 
\noindent 
is the Hamiltonian. Here $\lambda_k$ is the beam-beam coupling 
parameter, defined according to \cite{tzenov} 
\begin{equation}\label{eq:bbparam} 
\lambda_k = {\frac {R r_e N_{3-k} 
\beta_{kq}^{\ast}} {\gamma_{k0} L_{(3-k)x}}} 
{\frac {1 + \beta_{k0} \beta_{(3-k)0}} 
{\beta_{k0}^2}} \approx {\frac {2 R r_e N_{3-k} 
\beta_{kq}^{\ast}} {\gamma_{k0} L_{(3-k)x}}}. 
\end{equation} 
\noindent 
Moreover, ${\left( k = 1, 2 \right)}$ labels the beam, 
$f_k (q, p; \theta)$ is the distribution function, $\theta$ is the 
azimuthal angle, and $\nu_k$ is the betatron frequency in vertical 
direction. In addition, $R$ is the mean machine radius, $r_e$ is the 
classical electron radius, $N_{1,2}$ is the total number of particles 
in either beam, $V_k (q; \theta)$ is the normalized beam-beam 
potential, $\beta_{kq}^{\ast}$ is the vertical beta-function at the 
interaction point, and $L_{kx}$ is the horizontal dimension of the 
beam ribbon \cite{ruth}. 

Our goal is to determine a canonical transformation such that the 
new Hamiltonian is time-independent. As a consequence, the stationary 
solution of the Vlasov equation (\ref{eq:mixvlasov}) is expressed 
as a function of the new Hamiltonian. Following the procedure 
outlined in the Sec. 2 we transform Eqs. (\ref{eq:vlasovk}) 
-- (\ref{eq:hamiltonian}) according to 
\begin{equation}\label{eq:vlasovs} 
{\left[ F_0^{(k)}, {\cal K}_k \right]} 
\equiv 0, 
\end{equation} 
\begin{equation}\label{eq:hamil} 
{\frac {\partial S_k} {\partial \theta}} + 
\epsilon H_k {\left( q, {\frac {\partial S_k} 
{\partial q}}; \theta \right)} = 
{\cal K}_k {\left( q, P \right)}, 
\end{equation} 
\begin{equation}\label{eq:poiss} 
{\frac {\partial^2 V_k} {\partial q^2}} = 
4 \pi \! \int \! dP 
{\frac {\partial^2 S_k} {\partial q \partial P}} 
F_0^{(3-k)} {\left( q, P \right)}, 
\end{equation} 
\noindent 
where $\epsilon$ is again a formal small parameter, which will be set 
equal to unity at the end of the calculation. 

The next step is to expand the quantities $S_k$, ${\cal K}_k$ and 
$V_k$ in a power series in $\epsilon$, analogous to Eqs. 
(\ref{eq:expandsf}) -- (\ref{eq:expandphi}), according to 
\begin{equation}\label{eq:expands} 
S_k = qP + \epsilon G_k^{(1)} + 
\epsilon^2 G_k^{(2)} + \epsilon^3 G_k^{(3)} + 
\dots, 
\end{equation} 
\begin{equation}\label{eq:expandk} 
{\cal K}_k = \epsilon {\cal K}_k^{(1)} + 
\epsilon^2 {\cal K}_k^{(2)} + 
\epsilon^3 {\cal K}_k^{(3)} + \dots, 
\end{equation} 
\begin{equation}\label{eq:expandv} 
V_k = {\widetilde{V}}_k + 
\epsilon V_k^{(1)} + \epsilon^2 V_k^{(2)} + 
\epsilon^3 V_k^{(3)} + \dots, 
\end{equation} 
\noindent 
where 
\begin{equation}\label{eq:poisstilde} 
{\frac {\partial^2 {\widetilde{V}}_k} 
{\partial q^2}} = 4 \pi \! \int \! dP 
F_0^{(3-k)} {\left( q, P \right)}. 
\end{equation} 
\noindent 
Substitution of the above expansions (\ref{eq:expands}) -- 
(\ref{eq:expandv}) into Eqs. (\ref{eq:hamil}) and (\ref{eq:poiss}) 
yields perturbation equations that can be solved successively order 
by order. The results to third order in $\epsilon$ are briefly 
summarized below. 

{\it First Order}: $O(\epsilon)$ 
\begin{equation}\label{eq:hamil1} 
{\cal K}_k^{(1)} {\left( q, P \right)} = 
{\frac {\nu_k} {2}} {\left( P^2 + q^2 \right)} + 
{\frac {\lambda_k} {2 \pi}} 
{\widetilde{V}}_k (q), 
\end{equation} 
\begin{equation}\label{eq:generate1} 
G_k^{(1)} {\left( q, P; \theta \right)} = 
{\frac {i \lambda_k} {2 \pi}} 
{\widetilde{V}}_k (q) 
\sum \limits_{n \neq 0} 
{\frac {1} {n}} 
\exp{\left( i n \theta \right)}, 
\end{equation} 
\begin{equation}\label{eq:potential1} 
V_k^{(1)} (q; \theta) \equiv 0. 
\end{equation} 

{\it Second Order}: $O(\epsilon^2)$ 
\begin{equation}\label{eq:hamil2} 
{\cal K}_k^{(2)} {\left( q, P \right)} 
\equiv 0, 
\end{equation} 
\begin{equation}\label{eq:generate2} 
G_k^{(2)} {\left( q, P; \theta \right)} = 
- {\frac {\lambda_k \nu_k} {2 \pi}} P 
{\widetilde{V}}'_k (q) 
\sum \limits_{n \neq 0} 
{\frac {1} {n^2}} 
\exp{\left( i n \theta \right)}, 
\end{equation} 
\begin{equation}\label{eq:potential2} 
V_k^{(2)} (q; \theta) = - 
{\frac {\lambda_k \nu_k} {2 \pi}} 
{\widetilde{V}}_k^{(2)} (q) 
\sum \limits_{n \neq 0} 
{\frac {1} {n^2}} 
\exp{\left( i n \theta \right)}, 
\end{equation} 
\noindent 
where 
\begin{equation}\label{eq:poisstilde2} 
{\frac {\partial^2 {\widetilde{V}}_k^{(2)}} 
{\partial q^2}} = 
4 \pi {\widetilde{V}}''_k (q) 
\! \int \! dP 
F_0^{(3-k)} {\left( q, P \right)}. 
\end{equation} 

{\it Third Order}: $O(\epsilon^3)$ In third order we are interested 
in the new Hamiltonian, which is of the form 
\begin{equation}\label{eq:hamil3} 
{\cal K}_k^{(3)} {\left( q, P \right)} = 
{\frac {\lambda_k^2 \nu_k} {4 \pi^2}} 
\zeta (2) {\left[ 
{\widetilde{V}}_k^{\prime 2} (q) - 2 
{\widetilde{V}}_k^{(2)} (q) \right]}, 
\end{equation} 
\noindent 
where $\zeta (z)$ is Riemann's zeta-function \cite{abramowitz} 
\begin{equation}\label{eq:zeta} 
\zeta (z) = \sum \limits_{n=1}^{\infty} 
{\frac {1} {n^z}}. 
\end{equation} 

The effectiveness of the Hamiltonian formalism developed in the present 
paper has been demonstrated in two particular examples. In the first 
example discussed in the previous section, the short-scale dynamics 
is contained in the external focusing force acting on the beam, while 
an essential feature of the coherent beam-beam interaction treated above 
is the relatively fast variation of the collective interaction between 
the two colliding beams. The simplicity in applying the Hamiltonian 
averaging technique is embedded in the use of mixed canonical variables. 
Instead of expanding the generating function and the new Hamiltonian in 
terms of the new canonical coordinates and momenta \cite{channell,qin} 
one has to simply solve the Hamilton-Jacobi equations order by order. It 
should be pointed out that the mixed variable Hamiltonian formalism can 
be used to derive amplitude equations, describing processes of formation 
of patterns and coherent structures in a number of plasma and beam 
systems in which collective processes are important.

\section{THE EQUILIBRIUM DISTRIBUTION FUNCTION}

Since the new Hamiltonian ${\cal K}_k$ is time-independent (by 
construction), the equilibrium distribution function $F_0^{(k)}$ 
[see Eq. (\ref{eq:vlasovs})] is a function of the new Hamiltonian 
\begin{equation}\label{eq:equilibri} 
F_0^{(k)} (q, P) = {\cal G}_k 
{\left( {\cal K}_k \right)}, 
\end{equation} 
\noindent 
where 
\begin{equation}\label{eq:equiham} 
{\cal K}_k (q, P) = 
{\frac {\nu_k} {2}} {\left( P^2 \! + 
q^2 \right)} + 
{\frac {\lambda_k} {2 \pi}} 
{\widetilde{V}}_k (q) + 
{\frac {\lambda_k^2 \nu_k} {4 \pi^2}} 
\zeta (2) {\left[ 
{\widetilde{V}}_k^{\prime 2} (q) - 
2 {\widetilde{V}}_k^{(2)} (q) \right]}. 
\end{equation} 
\noindent 
Integrating Eq. (\ref{eq:equilibri}) over $P$ we obtain a nonlinear 
integral equation of the Haissinski type \cite{haissinski} for the 
equilibrium beam density profile $\varrho_0^{(k)}$ 
\begin{equation}\label{eq:density} 
\varrho_0^{(k)} (q) = \! \int \! dP 
{\cal G}_k {\left( {\cal K}_k \right)}, 
\end{equation} 
\noindent 
where 
\begin{equation}\label{eq:equihami} 
{\cal K}_k (q, P) = 
{\frac {\nu_k} {2}} {\left( P^2 \! + 
q^2 \right)} + \lambda_k \! \int \! 
dq' {\left| q - q' \right|}
\varrho_0^{(3-k)} {\left( q' \right)} 
+ 2 \lambda_k^2 \nu_k \zeta (2) 
{\cal F}_k (q), 
\end{equation} 
\begin{equation}\label{eq:calf} 
{\cal F}_k (q) = \int dq' dq'' {\cal Z} 
{\left( q-q', q'-q'' \right)} 
\varrho_0^{(3-k)} {\left( q' \right)} 
\varrho_0^{(3-k)} {\left( q'' \right)},  
\end{equation} 
\begin{equation}\label{eq:imped} 
{\cal Z} (u, v) = {\rm sgn} (u) 
{\rm sgn} (v) - 2 |u| \delta (v). 
\end{equation} 
\noindent 
Here ${\rm sgn} (z)$ is the well-known sign-function. 

Let us further specify the function 
${\cal G}_k {\left( {\cal K}_k \right)}$ and assume that it is given 
by the thermal equilibrium distribution 
\cite{davidson,tzenovbook,klimontovich} 
\begin{equation}\label{eq:gibbs} 
{\cal G}_k {\left( {\cal K}_k \right)} = 
{\cal N}_k \exp {\left( - {\frac {{\cal K}_k} 
{\varepsilon_k}} \right)}, 
\end{equation} 
\noindent 
where ${\cal N}_k$ is a normalization constant, defined according to 
\begin{equation}\label{eq:statsum} 
{\frac {1} {{\cal N}_k}} = \int dq dP \exp 
{\left[ - {\frac {{\cal K}_k {\left( q, P \right)}} 
{\varepsilon_k}} \right]}, 
\end{equation} 
\noindent 
and $\varepsilon_k$ is the unnormalized beam emittance. The second term 
in the Hamiltonian (\ref{eq:equihami}) can be transformed according to 
\begin{eqnarray}\nonumber 
\int \limits_{- \infty}^{\infty} \! dq' 
{\left| q - q' \right|}
\varrho_0^{(3-k)} {\left( q' \right)} = 
q - {\left \langle q_{3-k} \right \rangle} + 
2 \int \limits_q^{\infty} \! dq' 
{\left( q' - q \right)} 
\varrho_0^{(3-k)} {\left( q' \right)} 
\end{eqnarray} 
\begin{equation}\label{eq:transf} 
= q - {\left \langle q_{3-k} \right \rangle} + 
2 \int \limits_0^{\infty} \! dq_1 q_1 
\varrho_0^{(3-k)} {\left( q_1 + q \right)}, 
\end{equation} 
\noindent 
where 
\begin{equation}\label{eq:nmoment} 
{\left \langle q_k^n \right \rangle} = 
\int \limits_{- \infty}^{\infty} \! dq q^n 
\varrho_0^{(k)} {\left( q \right)}. 
\end{equation} 
\noindent 
Expanding the beam density $\varrho_0^{(3-k)} {\left( q_1 + q \right)}$ 
occurring in the integral in Eq. (\ref{eq:transf}) in a Taylor series 
and integrating by parts, we obtain 
\begin{equation}\label{eq:final} 
\int \limits_{- \infty}^{\infty} \! dq' 
{\left| q - q' \right|}
\varrho_0^{(3-k)} {\left( q' \right)} = 
{\left \langle q_{3-k}^{(+)} \right \rangle} - 
{\left \langle q_{3-k}^{(-)} \right \rangle} + 
{\left( 1 - 2 {\cal A}_{3-k} \right)} q + 
2 \sum \limits_{n=2}^{\infty} {\frac 
{{\cal C}_{3-k}^{(n)}} {n!}} q^n, 
\end{equation} 
\noindent 
where 
\begin{equation}\label{eq:meanqpm} 
{\left \langle q_k^{(+)} \right \rangle} = 
\int \limits_0^{\infty} \! dq q 
\varrho_0^{(k)} {\left( q \right)}, 
\qquad \qquad 
{\left \langle q_k^{(-)} \right \rangle} = 
\int \limits_{- \infty}^0 \! dq q 
\varrho_0^{(k)} {\left( q \right)}, 
\end{equation} 
\begin{equation}\label{eq:coeff} 
{\cal A}_k = \int \limits_0^{\infty} \! dq 
\varrho_0^{(k)} {\left( q \right)}, 
\qquad \qquad 
{\cal C}_k^{(n)} = {\left. 
{\frac {\partial^{n-2} \varrho_0^{(k)} 
{\left( q \right)}} {\partial q^{n-2}}} 
\right|}_{q=0}. 
\end{equation} 
\noindent 
Substituting (\ref{eq:gibbs}) and (\ref{eq:final}) into Eq. 
(\ref{eq:density}) we obtain 
\begin{equation}\label{eq:densit} 
\varrho_0^{(k)} {\left( q \right)} = {\cal N}_k 
{\sqrt{\frac {2 \pi \varepsilon_k} {\nu_k}}} \exp 
{\left[ - {\frac {\nu_k q^2} {2 \varepsilon_k}} - 
{\frac {\lambda_k} {\varepsilon_k}} {\left( 
1 - 2 {\cal A}_{3-k} \right)} q - 
{\frac {2 \lambda_k} {\varepsilon_k}} 
\sum \limits_{n=2}^{\infty} 
{\frac {{\cal C}_{3-k}^{(n)}} {n!}} q^n 
\right]}. 
\end{equation} 
\noindent 
Taking into account that 
\begin{equation}\label{eq:hermite} 
{\cal A}_k = {\frac {\pi \varepsilon_k} 
{\nu_k}} {\cal N}_k + O {\left( \lambda_k 
\right)}, \qquad \quad 
{\cal C}_k^{(n)} = {\cal N}_k 
{\sqrt{\frac {2 \pi \varepsilon_k} {\nu_k}}} 
(-1)^{n-2} 
{\left( {\frac {\nu_k} {2 \varepsilon_k}} 
\right)}^{(n-2)/2} H_{n-2} (0) + O 
{\left( \lambda_k \right)}, 
\end{equation} 
\noindent 
where $H_n (z)$ is the Hermite polynomial \cite{abramowitz} of order 
$n$, we obtain 
\begin{equation}\label{eq:dens} 
\varrho_0^{(k)} {\left( q \right)} = {\cal N}_k 
{\sqrt{\frac {2 \pi \varepsilon_k} {\nu_k}}} \exp 
{\left[ h_k {\left( q \right)} \right]}, 
\end{equation} 
\noindent 
where 
\begin{equation}\label{eq:lapfunc} 
h_k {\left( q \right)} = - {\frac {\nu_k q^2} 
{2 \varepsilon_k}} - {\frac {\lambda_k 
{\cal B}_{3-k} q} {\varepsilon_k}} - 
{\frac {2 \pi \lambda_k \varepsilon_{3-k} 
{\cal N}_{3-k}} {\varepsilon_k \nu_{3-k}}} 
{\left[ q \Phi {\left( a_{3-k} q \right)} + 
{\frac {e^{- a_{3-k}^2 q^2 }} {a_{3-k} 
{\sqrt{\pi}}}} \right]}, 
\end{equation} 
\noindent 
and 
\begin{equation}\label{eq:coefak} 
{\cal B}_k = 1 - {\frac {2 \pi \varepsilon_k 
{\cal N}_k} {\nu_k}}, 
\qquad \qquad \qquad 
a_k^2 = {\frac {\nu_k} {2 \varepsilon_k}}. 
\end{equation} 
\noindent 
Here, $\Phi(z)$ is the error function \cite{abramowitz}. 

In order to determine the normalization constant(s), ${\cal N}_k$, we 
utilize the method of Laplace to take the integral of the beam density 
$\varrho_0^{(k)} {\left( q \right)}$ over $q$. The first step consists 
in finding the extremum value(s) $q_k^{(e)}$ of the function(s) 
$h_k {\left( q \right)}$. They satisfy the (two) equation(s)
\begin{equation}\label{eq:extremum} 
{\frac {\nu_k q_k^{(e)}} {\varepsilon_k}} + 
{\frac {\lambda_k {\cal B}_{3-k}} 
{\varepsilon_k}} + 
{\frac {2 \pi \lambda_k \varepsilon_{3-k} 
{\cal N}_{3-k}} {\varepsilon_k \nu_{3-k}}} 
\Phi {\left( a_{3-k} q_k^{(e)} \right)} = 0. 
\end{equation} 
\noindent 
These are evidently maxima, since 
\begin{equation}\label{eq:maxima} 
h_k^{\prime \prime} {\left( q_k^{(e)} \right)} = 
- {\frac {\nu_k} {\varepsilon_k}} - 
{\frac {4 \pi \lambda_k \varepsilon_{3-k} 
{\cal N}_{3-k}} {\varepsilon_k \nu_{3-k}}} 
{\frac {a_{3-k}} {\sqrt{\pi}}} 
e^{- a_{3-k}^2 q_k^{(e)2}} < 0. 
\end{equation} 
\noindent 
Integrating the beam density (\ref{eq:dens}) over $q$, we obtain 
\cite{nayfeh}
\begin{equation}\label{eq:normalize} 
1 = 2 \pi {\cal N}_k {\sqrt{\frac {\varepsilon_k} 
{\nu_k {\left| h_k^{\prime \prime} 
{\left( q_k^{(e)} \right)} \right|}}}} 
\exp {\left[ h_k {\left( q_k^{(e)} 
\right)} \right]}. 
\end{equation} 
\noindent 
Equation (\ref{eq:normalize}) represents two transcendental equations 
for determining the normalization constants ${\cal N}_k$. For the beam 
centroid and the beam size, i.e., the first and the second moments of 
the beam density (\ref{eq:dens}), we obtain 
\begin{equation}\label{eq:centroid} 
{\left \langle q_k \right \rangle} = 
q_k^{(e)} + {\frac {2 {\cal N}_k} 
{\left| h_k^{\prime \prime} 
{\left( q_k^{(e)} \right)} \right|}} 
{\sqrt{\frac {2 \pi \varepsilon_k} 
{\nu_k}}} \exp {\left[ h_k {\left( 
q_k^{(e)} \right)} \right]}, 
\end{equation} 
\begin{equation}\label{eq:size} 
{\left \langle q_k^2 \right \rangle} = 
q_k^{(e)2} + {\frac {4 {\cal N}_k 
q_k^{(e)}} {\left| h_k^{\prime \prime} 
{\left( q_k^{(e)} \right)} \right|}} 
{\sqrt{\frac {2 \pi \varepsilon_k} 
{\nu_k}}} \exp {\left[ h_k {\left( 
q_k^{(e)} \right)} \right]} + 
2 \pi {\cal N}_k {\sqrt{\frac {\varepsilon_k} 
{\nu_k}}} {\left[ {\frac {1} 
{\left| h_k^{\prime \prime} 
{\left( q_k^{(e)} \right)} \right|}} 
\right]}^{3/2} \exp {\left[ h_k {\left( 
q_k^{(e)} \right)} \right]}. 
\end{equation} 

In order to proceed further, we assume that the beam-beam coupling 
parameter $\lambda_k$ is small, and expand the equilibrium beam density 
$\varrho_0^{(k)} {\left( q \right)}$ in a perturbation series in 
$\lambda_k$ according to 
\begin{equation}\label{eq:perturb} 
\varrho_0^{(k)} {\left( q \right)} = 
\varrho_{00}^{(k)} {\left( q \right)} + 
\lambda_k \varrho_{01}^{(k)} {\left( q \right)} + 
\dots, 
\end{equation} 
\noindent 
where 
\begin{equation}\label{eq:densit0} 
\varrho_{00}^{(k)} {\left( q \right)} = 
{\frac {{\cal N}_k {\sqrt{\pi}}} {a_k}} 
\exp {\left( - a_k^2 q^2 \right)}, 
\end{equation} 
\noindent 
and 
\begin{equation}\label{eq:densit1} 
\varrho_{01}^{(k)} {\left( q \right)} = - 
{\frac {1} {\varepsilon_k}} 
{\left\{ {\cal B}_{3-k} q + 
{\frac {\pi {\cal N}_{3-k}} {a_{3-k}^2}}
{\left[ q \Phi {\left( a_{3-k} q \right)} + 
{\frac {e^{- a_{3-k}^2 q^2 }} {a_{3-k} 
{\sqrt{\pi}}}} \right]} \right\}} 
\varrho_{00}^{(k)} {\left( q \right)}. 
\end{equation} 
\noindent 
The main goal in what follows is to determine the normalization 
constant(s) ${\cal N}_{k0}$. To do so we integrate Eq. 
(\ref{eq:perturb}) over $q$.  As a result of simple algebraic 
manipulations, we obtain 
\begin{equation}\label{eq:normbas} 
{\frac {\pi {\cal N}_k} {a_k^2}} - 
{\frac {\pi {\sqrt{\pi}} \lambda_k} 
{\varepsilon_k a_k^3 a_{3-k}^3}} 
{\sqrt{ a_k^2 + a_{3-k}^2}} 
{\cal N}_k {\cal N}_{3-k} = 1. 
\end{equation} 

Introducing the new unknowns 
\begin{equation}\label{eq:newunknow} 
{\cal M}_k = {\frac {\pi {\cal N}_k} 
{a_k^2}}, 
\end{equation} 
\noindent 
we can write the two equations for determining ${\cal M}_{1,2}$ as 
\begin{eqnarray}\label{eq:quadr} 
1 = {\cal M}_1 - b_1 {\cal M}_1 {\cal M}_2, 
\nonumber \\ 
\\ 
1 = {\cal M}_2 - b_2 {\cal M}_1 {\cal M}_2, 
\nonumber 
\end{eqnarray} 
\noindent 
where 
\begin{equation}\label{eq:coeffib} 
b_1 = {\frac {\lambda_1} 
{\varepsilon_1 {\sqrt{\pi}}}} 
{\frac {\sqrt{a_1^2 + a_2^2}} {a_1 a_2}}, 
\qquad \qquad \qquad 
b_2 = {\frac {\lambda_2} 
{\varepsilon_2 {\sqrt{\pi}}}} 
{\frac {\sqrt{a_1^2 + a_2^2}} {a_1 a_2}}. 
\end{equation} 
\noindent 
From Eq. (\ref{eq:quadr}), as a result of simple algebraic manipulations 
we obtain the quadratic equation
\begin{equation}\label{eq:quadratic} 
b_2 {\cal M}_1^2 - {\left( b_2 -b_1 + 1 \right)} 
{\cal M}_1 + 1 = 0 
\end{equation} 
\noindent 
for ${\cal M}_1$, and the equation 
\begin{equation}\label{eq:linear} 
b_1 {\cal M}_2 = b_2 {\cal M}_1 + b_1 - b_2 
\end{equation} 
\noindent 
for determining ${\cal M}_2$ once ${\cal M}_1$ is known. Equation 
(\ref{eq:quadratic}) has one real double root if and only if the 
discriminant 
\begin{equation}\label{eq:discrim} 
{\cal D} = {\left( b_2 -b_1 + 1 
\right)}^2 - 4 b_2 
\end{equation} 
\noindent 
is equal to zero. This gives 
\begin{equation}\label{eq:soldiscr} 
b_2 = {\left( {\sqrt{b_1}} \pm 1 \right)}^2. 
\end{equation} 
\noindent 
Since the scaled normalization constants ${\cal M}_{1,2}$ should be 
positive we choose 
\begin{equation}\label{eq:soldiscrim} 
b_2 = {\left( {\sqrt{b_1}} - 1 \right)}^2. 
\end{equation} 
\noindent 
Thus we obtain 
\begin{equation}\label{eq:normconst} 
{\cal M}_1 = {\frac {1} {\sqrt{b_2}}} = 
{\frac {1} {\left| {\sqrt{b_1}} - 1 \right|}}, 
\qquad \qquad \qquad 
{\cal M}_2 = {\frac {1} {\sqrt{b_1}}}. 
\end{equation} 

To conclude this section we note that in the case of ${\cal D} \neq 0$ 
we have two solutions for either ${\cal M}$, i.e., 
\begin{equation}\label{eq:flipflopsol} 
{\cal M}_1^{(1,2)} = {\frac {b_2 - b_1 + 
1 \pm {\sqrt{\cal D}}} {2 b_2}}, 
\qquad \qquad \qquad 
{\cal M}_2^{(1,2)} = {\frac {b_1 - b_2 + 
1 \pm {\sqrt{\cal D}}} {2 b_1}}. 
\end{equation} 
\noindent 
Note also that the discriminant ${\cal D}$ is invariant (does not 
change) under permutation of $b_1$ and $b_2$. In other words, four 
different physically realizable situations are possible for a wide 
range of parameters 
\begin{equation}\label{eq:flipflopar} 
0 < b_2 < 1 + b_1. 
\end{equation} 
\noindent 
The inequality in Eq. (\ref{eq:flipflopar}) has been obtained under 
the condition that both solutions in Eq. (\ref{eq:flipflopsol}) are 
positive. This case corresponds to the so-called ``flip-flop'' state 
\cite{otboyev} of the two colliding beams, which is a bifurcated state 
better to be avoided.

\section{CONCLUSIONS}

We have developed a systematic canonical perturbation approach 
that removes rapidly oscillating terms in Hamiltonians of quite 
general form. The essential feature of this approach is the use 
of mixed canonical variables. For this purpose the Vlasov-Poisson 
equations are transformed to mixed canonical variables, and an 
appropriate perturbation scheme is chosen to obtain the equilibrium 
phase space density. It is worthwhile to note that the perturbation 
expansion outlined in the preceding section can be carried out to 
arbitrary order, although higher-order calculations become very 
tedious. 

In conclusion, it is evident from the present analysis that the 
Hamiltonian averaging technique developed here represents a powerful 
formalism with applications ranging from beam propagation through a 
periodic focusing lattice (Sec. 3) to coherent beam-beam interaction 
(Secs. 4 and 5). For example, in the application to the coherent 
beam-beam interaction, the rapidly oscillating terms due to the 
periodic beam-beam kicks have been averaged away, leading to a new 
time-independent Hamiltonian (Sec. 4). Furthermore, the equilibrium 
distribution functions have been obtained as a general function of 
the new Hamiltonian, and coupled set of integral equations for the 
beam densities has been obtained (Sec. 5). An intriguing feature of 
the analysis in Sec. 5 is the derivation of a condition for existence 
of the so-called ``flip-flop'' state \cite{otboyev} of the two 
colliding beams, which is a bifurcated state better to be avoided in 
experimental applications. 

We reiterate that the formalism developed here is strictly speaking 
non-canonical and in general does not provide complete elimination of 
fast oscillating terms in the transformed Vlasov equation in the mixed 
variables. Nevertheless, such an elimination can be performed in the 
new Hamiltonian in the mixed variables. Moreover, if the distribution 
function is assumed to be an arbitrary function of the new 
time-independent Hamiltonian, it is a stationary solution of the 
nonlinear Vlasov equation in the mixed variables. 

Finally, we reiterate that the mixed variable Hamiltonian formalism 
developed in the present analysis can be used to derive amplitude 
equations, describing processes of formation of patterns and coherent 
structures in a number of plasma and beam systems in which collective 
processes are important. 

\section{ACKNOWLEDGMENTS}

We are indebted to S.A. Heifets for many fruitful discussions 
concerning the subject of the present paper. It is a pleasure to thank 
H. Qin for illuminating discussions and comments. This research was 
supported by the U.S. Department of Energy.

\end{document}